\documentclass{ws-procs9x6-cpt16}

\begin{document}

\newcommand{\refeq}[1]{(\ref{#1})}
\def\etal {{\it et al.}}

\title{Searches for Exotic Interactions with the nEDM Experiment}

\author{V.\ Bondar\footnote{Swiss Government Excellence Scholar on leave of absence from  National Technical University of Ukraine ``Kyiv Polytechnic Institute''}}

\address{Paul Scherrer Institut,
Villigen PSI, CH-5232, Switzerland}

\author{On behalf of the nEDM at PSI Collaboration (www.psi.ch/nedm)}

\begin{abstract}
Ultracold neutrons were used to search for signals indicating the violation of Lorentz and CPT invariance or the existence of dark matter using the spectrometer to search for an electric dipole moment of the neutron.
\end{abstract}

\bodymatter

\section{Introduction}
Lorentz and CPT invariance are cornerstones of the foundation of modern physics and essential as fundamental hypothesis of the Standard Model (SM) of particle physics.
The Standard-Model Extension including Lorentz- and CPT-violating terms has been presented in Ref.\ \refcite{Co97}.
An anisotropic cosmic background field coupled to spin-1/2 particles would violate Lorentz invariance~(LI), which can be tested using spin precession of ultracold neutrons~(UCN).
Going beyond the SM, searches for dark matter (DM) are motivated by cosmological observations.\cite{Ba15} A possible DM candidate is a mirror matter,\cite{Be06} which is an SM copy. It  interacts with SM particles via gravity and presents a viable explanation to DM. Another class of DM candidates are axion-like particles arising from the Peccei-Quinn theory,\cite{Mo84} which are possible mediators of short-range spin-dependent forces.

Spectrometers dedicated to searches for a neutron electric dipole moment (nEDM) using UCN also lend themselves in an excellent way to searches for signals indicating LI, CPT violation and other beyond-SM interactions. Here we report on results obtained using nEDM setup to set a limit on mirror neutron oscillations,\cite{Ba07}
UCN coupling to LI violating cosmic background,\cite{Al09} and
short-range spin-dependent forces.\cite{Af15}

\section{Ultracold neutrons and the nEDM spectrometer}

The measurements were performed with the apparatus which was previously used to obtain the current upper limit on the nEDM, $d_{n}~<~3.0~\times ~10^{-26}~e\textrm{cm}$, at the Institute Laue-Langevin\cite{nEDM1,nEDM2} and is now located at the UCN source of the Paul Scherrer Institute.\cite{source}
The  UCNs of energies below 160\,neV are confined  in a  chamber in vacuum exposed to  a homogenous magnetic field of $B\approx 1$ $\mu$T.
After filling the chamber with $N_0$   UCNs, the remaining UCN number after a storage time $t$ is given  by: 
\begin{equation}
\label{storage}
N(t)= N_{0}  e^{- t/\tau} = N_{0} e^{-(\Gamma_{\beta}+\Gamma_{\rm loss}+\Gamma_{\rm ex}) t},
\end{equation}

\noindent   where the decay time constant $1/\tau = \sum{\Gamma_i}$ includes the neutron lifetime $\tau_{\beta} =1/\Gamma_\beta$ and $\Gamma_{\rm loss}$ takes into account  all other neutron-loss mechanisms. For simplicity, here, we  neglect energy dependencies of the losses. An exotic interaction would introduce an additional channel $\Gamma_{\rm ex}$  of UCN loss.

The spectrometer was designed to measure tiny changes of the Larmor precession frequency $\omega_{\rm n}=\gamma_{\rm n}|B|$ using Ramsey's method of separated oscillatory fields. 
Possible exotic interaction coupling the spin of UCN to a pseudomagnetic field will shift the Larmor precession $\omega_{\rm n}$.  The measurement of the Larmor precession $\omega_{\rm Hg}$ of polarized $^{\text{199}}$Hg atoms in the same volume allows to correct for magnetic field fluctuations. 
A statistical sensitivity and an excellent control over systematic effects was reported in Ref.\ \refcite{SE}.

\section{Experimental measurements and results}

We searched for neutron--mirror-neutron (nn$'$) oscillations by comparing the UCN storage in the presence and the absence of a magnetic field.\cite{Ba07}
The transitions into mirror neutrons would add a neutron-loss channel and the storage time constant measured with magnetic field would be longer than without magnetic field.
Measuring neutron-loss rates in different magnetic fields, we set  a limit on the nn$'$ oscillation time $\tau_{\rm nn^{\prime}}  > 103 \, \textrm{s  (95\% C.L.)}$.
Later, a more stringent limit\cite{Se08} of  $\tau_{\rm nn^{\prime}}  > 448 \, \textrm{s  (90\% C.L.)}$ was reported.
Further studies take into account mirror-magnetic fields\cite{Mirr09} and the latest analysis\cite{Be12}  suggests  more experiments  to be performed.

In order to test LI, we searched for a sidereal modulation of the neutron Larmor frequency induced by $b_{\bot}$, the component of \textbf {b} orthogonal to the Earth's rotation axis.  While the Earth is rotating, the local magnetic field rotates with the Earth, and the interaction  appears in a harmonic change of  $R=\omega_{\rm n}/\omega{\rm_{Hg}}$.
No daily variations of $R$ were found and an upper limit on the LI violating cosmic background $ b_{ \bot } =1\,\times\,10^{-20}$ eV (95\% C.L.) was deduced.
Further, the nEDM magnitude modulation in periods of 12 and 24~hours was constrained\cite{Al10} to $ d_{12} < 15\times10^{-25}~e$cm and $ d_{24} <10\times10^{-25}~e$cm.

In a search for spin-dependent interactions,  the coupling between polarized UCN and unpolarized nucleons of the chamber's surface was studied.\cite{Af15}
The interaction strength can be described by a pseudomagnetic field  and is  proportional to the CP violating product of scalar and pseudoscalar coupling constants,  $ g_{s}  g_{p} $.
No change in the Larmor precession frequency induced by a pseudomagnetic interaction was found, which results in the most stringent upper limit for neutrons of $ g_{s}g_{p} \lambda^{2}<2.2\times10^{-27}$ m$^2$ for a range of 
1 $\mu$m $< \lambda < $ 5 mm.
An improved limit\cite{Gu15} of  $g_{s}g_{p}\lambda^{2}=2.6\times10^{-28}$~m$^2$  was set with  $^3$He.
The sensitivity of the spectrometer  to  spin-dependent forces can be increased by replacing one of the surfaces with a material of higher nucleon density.
A recently developed spin-echo technique will allow to take into account systematic effects from the interplay of a vertical magnetic-field gradient and the gravitational UCN density striation.\cite{SE}

In conclusion,
a significantly improved sensitivity is expected in the near future with the the same  experimental installation due to 
increase of UCN number detected after storage and better control over systematic effects.

\section*{Acknowledgements}
I would like to thank greatly the organizers of the CPT'16 meeting, the UCN group at PSI, and colleagues from the nEDM collaboration.

\end{document}